\documentclass[a4,11pt]{article}

\usepackage{graphicx}
\usepackage{authblk} %for authors with multiple affiliations

\usepackage{amsthm}
\usepackage{float}
\restylefloat{table}
\usepackage{amsmath}
\usepackage{amssymb}
\usepackage{amsfonts,cancel}
\usepackage[usenames]{color}
\usepackage{authblk} %for authors with multiple affiliations
\usepackage{soul} % enables one to strike out text 
\usepackage{appendix}
\usepackage{longtable}

\newcommand{\be}{\begin{equation}}
	\newcommand{\ee}{\end{equation}}
\newcommand{\bea}{\begin{eqnarray}}
	\newcommand{\eea}{\end{eqnarray}}
\newcommand{\Kdt}{{\hbox{\tiny K}}}
\newcommand{\rovno}{\!\!\!& = &\!\!\!} % rovnitko se zarovnanim pro pouztiti v eqnarray

\def \d {{\rm d}}
\def \dd {{\rm d}}

\def \pul {\textstyle{\frac{1}{2}}}

\def \H {\mathcal{H}}

\def \B {\mathcal{B}}

\newcommand{\beqn}{\begin{eqnarray}}
	\newcommand{\eeqn}{\end{eqnarray}}

\usepackage{amsmath,amsthm,amssymb,cite,enumerate}

\def \d {{\rm d}}

\usepackage{amsmath}

\begin{document}

	\title{Charged black holes in  quadratic gravity}
	
	\author[1]{Vojt\v ech Pravda\thanks{pravda.math.cas.cz}}
	
	\author[1]{Alena Pravdov\'a \thanks{pravdova.math.cas.cz}}

	\author[1,2]{George Turner \thanks{turner.math.cas.cz}}

 \affil[1]{Institute of Mathematics of the Czech Academy of Sciences, \newline \v Zitn\' a 25, 115 67 Prague 1, Czech Republic}
 
\affil[2]{ Charles University, V~Hole\v{s}ovi\v{c}k\'ach~2, 180~00 Prague 8, Czech Republic}

 %\author{Vojt\v ech Pravda$^\diamond$, Alena Pravdov\' a$^\diamond$, George Turner^\star$
%	\\
%	\vspace{0.05cm} \\
%	\vspace{0.05cm} \\
%	{\small $^\diamond$ Institute of Mathematics, Academy of Sciences of the Czech Republic}, \\
%	{\small \v Zitn\' a 25, 115 67 Prague 1, Czech Republic} \\
%	{\small $^\star$ Institute of Theoretical Physics, Faculty of Mathematics and Physics,} \\
%	{\small Charles University, V~Hole\v{s}ovi\v{c}k\'ach~2, 180~00 Prague 8, Czech Republic
%	} \\
%{\small  E-mail: %\texttt{pravda@math.cas.cz, %pravdova@math.cas.cz,turner.math.cas.cz}}\\
%	\\}

	\maketitle

	\begin{abstract}
		We study electrically charged, static, spherically symmetric black holes in quadratic gravity using the conformal-to-Kundt technique, which leads to a considerable simplification of the
field equations. We study the solutions using a
Frobenius-like approach of power-series expansions. The indicial equations restrict the set of possible leading powers to a few cases,
describing, e.g., black holes, wormholes, or naked singularities.

We focus on the black hole case and derive recurrent formulas for all series coefficients of the infinite power-series
expansion around the horizon. The solution is characterized by electric charge $q$, the black-hole radius $a_0$, and the Bach parameter $b$ related to the strength of the Bach tensor at the horizon. However, the Bach parameter has to be fine-tuned to ensure asymptotic flatness. The fine-tuning of $b$ for a given $q$ and $a_0$ returns up to two values, describing two branches of asymptotically flat, static, spherically symmetric, charged black holes in quadratic gravity. This is in agreement with previous numerical works.

We discuss various physical properties of these black holes, such as their asymptotic mass, temperature, photon spheres, and black-hole shadows. A straightforward generalization to dyonic black holes in quadratic gravity is also briefly mentioned.

	\end{abstract}

\section{Introduction}

General relativity is currently our best theory of gravity, predicting and describing such fundamental physical phenomena as gravitational waves, cosmic expansion, and black holes. This classical theory of gravity, however, does not address quantum effects. From the effective field-theory view, higher-order correction terms should be added to the Einstein-Hilbert action. In this paper, we focus on quadratic gravity, where
 correction terms quadratic in the curvature are present.

Recently there has been considerable interest in spherically symmetric black holes in quadratic gravity, following early results of \cite{Stelle77,Stelle78}.     It is well known that in four dimensions, Einstein spaces obey the vacuum field equations of quadratic gravity identically \cite{Buchdahl48_2,Buchdahl48_3}. The Schwarzschild black hole is, therefore, clearly a vacuum solution to quadratic gravity. Recently, however, it has been shown that quadratic gravity also admits another static, spherically symmetric black hole solution over and above Schwarzschild \cite{Luetal15,Luetal15b}, violating the Birkhoff theorem of standard general relativity.

%results which were generalised to include non-vanishing cosmological constant $\Lambda$ in \cite{Svarcetal18,Pravdaetal21}.  
	
	   These non-Schwarzschild (or Schwarzschild-Bach) black holes admit one additional parameter - the Bach parameter $b$. % In the case with vanishing $\Lambda$, 
    However, the Bach parameter $b$ has to be fine-tuned to ensure asymptotic flatness \cite{Luetal15,Luetal15b,Luetal17,Podolskyetal18}, leading effectively to a one-parameter family of asymptotically flat Schwarzschild-Bach black holes. In contrast, in the case of static, spherically symmetric black holes with non-vanishing cosmological constant $\Lambda$ \cite{Svarcetal18,Pravdaetal21}, fine-tuning of the Bach parameter is not necessary since the Schwarzschild-Bach-(A)dS black holes are asymptotically (A)dS within certain continuous ranges of parameters, cf. \cite{Luetal12,PraPraOrt23}.

Recent works also studied the stability of static, spherically symmetric black holes in quadratic gravity.	Long-wavelength instability of Schwarzschild-Bach black holes %complementing Gregory-Laflamme instability of the Schwarzschild branch
	 has been found \cite{Held23} for horizon radii smaller than a critical value. Thus there seems to be a lower bound for the horizon radius of stable Schwarzschild-Bach black holes in quadratic gravity \cite{Held23}.

Charged, static, spherically symmetric black holes in quadratic gravity have been studied in \cite{Lin2017,Wu20} using numerical methods. In this paper, we will study these black holes using the conformal-to-Kundt technique 
recently employed in  \cite{Podolskyetal18,Podolskyetal20,Svarcetal18,Pravdaetal21,PraPraOrt23}. This approach leads to a significant simplification of the field equations.  This will enable us
to find recurrent formulas for coefficients of power-series expansions of the metric
functions and apply analytical techniques to the study of these black holes.

 In Section \ref{sec_background}, we revisit the electro-vacuum field equations of quadratic gravity and the conformal-to-Kundt approach.
 
 In section \ref{sec_FEqs}, we derive the field equations in the Kundt coordinates. In passing, we also mention a straightforward generalization of the electrically charged black hole studied in this paper to the case of a dyonic black hole in quadratic gravity. 
	
	In section \ref{sec_expr0}, we employ a Frobenius-like approach to solve the field equations in the Kundt coordinates in the vicinity of a generic hypersurface of constant radius.  The field equations impose constraints on the dominant powers in the expansions, leading to a few allowed classes of static, spherically symmetric power-series solutions of quadratic gravity with electromagnetism. Besides black holes, these solutions may also represent wormholes and naked singularities. However, in this paper, we focus on the black-hole solutions.

In section \ref{sec_non-extremal}, we concentrate on solutions admitting nonextremal horizons. We derive recurrent formulas for coefficients of power-series expansions of the metric. We observe that (at least for a certain range of parameters), the coefficients asymptotically approach a geometric series, which allows us to straightforwardly estimate the radii of convergence of the solutions.

Let us stress that in contrast with the vacuum case, where the Schwarzschild black hole solves the vacuum field equations of quadratic gravity, in the electrovacuum case, the Reissner-Nordstr\" om black hole does \emph{not} obey the electrovacuum field equations. Instead, in agreement with the numerical results of \cite{Lin2017,Wu20}, we identify and study two branches of spherically symmetric, charged black holes. One represents charged Schwarzschild black holes in quadratic gravity (distinct from Reissner-Nordstr\" om), and the other one represents charged Schwarzschild-Bach black holes.
In vacuum, the Schwarzschild and Schwarzschild-Bach black hole families intersect for a critical radius (e.g. $\bar r_0 \sim 0.876$, for parameters chosen in \cite{Luetal15b}). In the charged case, this is not always true, at least for a sufficiently large charge.

   In section \ref{sec_non-extremal}, we also study photon spheres and black-hole shadows of these charged black holes, benefiting from the simplification of photon-sphere and black-hole shadow description in the Kundt coordinates (as recently pointed out in \cite{Ortaggio24prep}).

In section \ref{sec_[0,2]}, we briefly discuss a case admitting an extremal horizon. Again, we provide recurrent formulas for coefficients of power-series expansions of the corresponding metric. However, it seems much more difficult to fine-tune this case for asymptotic flatness. Therefore from our work, the existence of a charged, extremal, asymptotically flat black hole is inconclusive and will require further study.\footnote{In the numerical work of \cite{Lin2017}, it is stated that extremal black holes in this theory could exist in the Schwarzschild branch while not in the Schwarzschild-Bach branch.  }

 Finally, for completeness, in the Appendix,  we list the fine-tuned values of the  Bach parameter for various values of charge and radius of the black hole.  
\section{Background}
\label{sec_background}

\subsection{The  quadratic gravity field equations}
\label{subsec_QG}

The action of quadratic gravity coupled with electromagnetism reads
\be
S {=\int \d^4 x\, \sqrt{-g}{\cal L}}= \int \d^4 x\, \sqrt{-g}\, 
\Big( \gamma \, R  +\beta\,R^2  - \alpha\, C_{abcd}\, C^{abcd}
-\frac{\kappa}{2} F_{ab} F^{ab}\Big)\,,
\label{action}
\ee
where ${\gamma=1/(16\pi G)}$, $G$ is the Newtonian constant (we will set $G=1=c$), %$\Lambda $ is the cosmological constant, 
and  $\alpha$, $\beta$  are {coupling} constants of quadratic gravity.

The above action \eqref{action} leads to the following gravitational field equations
\be
\gamma \left(R_{ab} - {\pul} R\, g_{ab}\right)-4 \alpha\,B_{ab}
+2\beta\left(R_{ab}-\tfrac{1}{4}R\, g_{ab}+ g_{ab}\, \Box - \nabla_b \nabla_a\right) R = \kappa T_{ab}\,, \label{fieldeqsEW}
\ee
where the energy-momentum tensor $T_{ab}$ reads
\be
T_{ab} = F_{ac} {F_{b}}^{ c} - \frac{1}{4} {g_{ab} F_{cd} F^{cd}}\label{elTab}
\ee
and  $B_{ab}$ is the {Bach tensor}
\be
B_{ab} \equiv \big( \nabla^c \nabla^d + {\pul} R^{cd} \big) C_{acbd} \ . \label{defBach}
\ee
The Bach tensor is traceless, symmetric, conserved, and well-behaved under conformal transformations of the metric tensor $g_{ab}=\Omega^2 \tilde g_{ab}$:
\begin{equation}
	g^{ab}B_{ab}=0 \,, \qquad B_{ab}=B_{ba} \,, \qquad
	\nabla^b B_{ab}=0
	\,, \qquad B_{ab}=\Omega^{-2}\tilde B_{ab}\,.
	\label{Bachproperties}
\end{equation}
Note that for Einstein spacetimes, the Bach tensor vanishes  \cite{Buchdahl53} and thus for these spacetimes, the vacuum field equations  (eq. \eqref{fieldeqsEW} with $T_{ab}=0$) hold identically, making Einstein spacetimes, such as the Schwarzschild metric,  trivial solutions of quadratic gravity as noted in Introduction. This ``immunity'' of Einstein spaces to the addition of quadratic gravity terms to the vacuum Einstein field equations, however, does not generalize to electrovacuum solutions of Einstein gravity. In particular, Reissner-Nordstr\" om black holes do not obey the quadratic gravity field equations with electromagnetism \eqref{fieldeqsEW}.  

In the vacuum case, the trace no-hair theorem of \cite{Nelson2010}, \cite{Luetal15b} states that for asymptotically flat, static, spherically symmetric black holes in quadratic gravity, the Ricci scalar $R$  vanishes throughout the spacetime.
Noting that the energy-momentum $T_{ab}$ of an electromagnetic field is also traceless, it has been argued in \cite{Lin2017} that the vanishing of $R$ applies also to the electrovacuum case. This will be thus assumed in the rest of this paper.

  The field equations \eqref{fieldeqsEW} then reduce significantly to 
\be
R_{ab}-4k\, B_{ab} = \kappa' T_{ab},
\label{eq:feq}
\ee
where 
\be
k\equiv\frac{\alpha}{\gamma}, \qquad  \kappa' \equiv \frac{\kappa}{\gamma} ,
\label{k}
\ee 
assuming  $\gamma\neq0$.

\subsection{Conformal-to-Kundt ansatz}

Instead of using the 
	standard spherically symmetric metric
\be
\dd s^2 = -h(\bar r)\,\dd t^2+\frac{\dd \bar r^2}{f(\bar r)}+ \bar r^2 \dd\omega^2\,,\ \ \dd\omega^2= \dd \theta^2 + \sin^2 \theta \dd \phi^2, \ \ \  \ 
 \label{physmet}
\ee	
we will employ its conformal-to-Kundt form ~\cite{Pravdaetal17,Podolskyetal18,Podolskyetal20}
\be
\dd s^2 \equiv \Omega^2(r) \,\dd s^2_\Kdt = \Omega^2(r)
\Big[\,\dd \omega^2 -2\,\dd u\,\dd r+\H(r)\,\dd u^2 \,\Big]\,,
\label{BHmetric}
\ee
which leads to a considerable simplification of the field equations.

This metric admits a gauge freedom
\be
r \to \lambda\,r+\upsilon\,, \qquad u \to \lambda^{-1}\,u \,, 
\label{scalingfreedom}
\ee
where $\lambda\,, \upsilon$ are constants.
In addition, the metric \eqref{physmet} admits also  a {time-scaling freedom} 
\be
t\to  t/\sigma \ \  \Rightarrow\ \ h\to h\,\sigma^2\,,
\label{scaling-t}
\ee
where the constant ${\sigma \ne 0}$ 
can be used to
adjust value of $h$ at a chosen radius ${\bar r}$.

The standard metric form \eqref{physmet} can be obtained (assuming $\Omega'\neq0\neq \H$) by \cite{Pravdaetal17}
\begin{equation}
	\bar{r} = \Omega(r)\,, \qquad t = u - \int\! \frac{\dd r}{\H(r)} \,,
	\label{to static}
\end{equation}
leading to
\begin{equation}
	h = -\Omega^2\, \H , \quad f = -\left(\frac{\Omega'}{\Omega}\right)^2 \H  , 
	\label{Schwarz}  
\end{equation}
where a prime denotes differentiation with respect to $r$.

The {Killing horizon}, where the Killing vector field ${\partial_u}={\partial_t}$ becomes null, is located 
 at the zeros $r=r_h$ of the metric function $\H $, 
\begin{equation}
\H \big|_{r=r_h}=0\,. \label{horizon}
\end{equation} 
Then using \eqref{Schwarz}, also $h({\bar r_h})=0=f({\bar r}_h)$, where ${\bar r}_h=\Omega(r_h)$.

Curvature invariants constructed from the Bach and Weyl tensors have the same form for vacuum \cite{Podolskyetal18} and charged case 
\bea
B_{ab}\, B^{ab} &=&  \tfrac{1}{72}\,\Omega^{-8}\,\big[(\B_1)^2 + 2(\B_1+\B_2)^2\big] \,,\label{invB}\\
C_{abcd}\, C^{abcd} &=&  \tfrac{1}{3}  \,\Omega^{-4}\,\big({\H}'' +2\big)^2 \,, \label{invC}
%\\
%{\tb{ R_{ab}R^{ab} }} &=&(4k)^2 B_{ab}B^{ab}
%+\kappa'^2 \left[F_{ac}{F_b}^cF^{ad}{F^b}_d
%-\tfrac{1}{4}(F_{ab}F^{ab})^2\right]
%+8k\kappa' B^{ab} F_{ac}{F_b}^c\,,
%\label{invR}
\eea
where
\bea
&& \B_1 \equiv {\H}{\H}''''\,, \label{B1}\\
&& \B_2 \equiv {\H}'{\H}'''-\tfrac{1}{2}{{\H}''}^2 +2\,. \label{B2}
\eea

Note that the Kundt coordinates are also useful for the description of conformally invariant properties of spacetimes, such as photon spheres, see sec. \ref{sec_photon}.  

%------------------------

\subsubsection{The Reissner-Nordstr\" om black hole }

Recall that the Schwarzschild metric is a vacuum solution of quadratic gravity.
In contrast, the Reissner-Nordstr\" om metric is not a solution of electrovacuum quadratic gravity \eqref{eq:feq}, \eqref{elTab}. Nevertheless,  it will be useful for comparison with actual solutions.

 In  the Kundt coordinates, the Reissner-Nordstr\" om  metric functions  read
\be
\Omega(r) = -\frac{1}{r},\ \ 
\H(r)= -r^2-2 M r^3- \frac{Q^2}{4 \pi\epsilon_0} r^4\,.\label{RNgen}
\ee

In the vicinity of a horizon,  which is located at a non-zero root
of  $\H(r_h)=0$,
%where $\H(r_h)=0=1 +2 M r_h+\frac{Q^2}{4 \pi\epsilon_0}r_h^2 =0$, 
eq. \eqref{RNgen} gives
\bea
\H(r)&=&
- \left(2+6 M r_h+\frac{Q^2}{ \pi\epsilon_0} r_h^2\right)r_h (r-r_h)
-\frac{1}{2}  \left(2+12 M r_h+3 \frac{Q^2}{ \pi\epsilon_0}r_h^2\right) (r-r_h)^2\nonumber\\
&&   - \left(2 M+ \frac{Q^2}{ \pi\epsilon_0} r_h\right)  (r-r_h)^3
      -\frac{Q^2}{ 4\pi\epsilon_0} (r-r_h)^4\,.
      \eea

\section{The field equations in Kundt coordinates}
\label{sec_FEqs}
\subsection{Maxwell equations}

In agreement with the previous works \cite{Lin2017,Wu20}, we start with the following ansatz for the electromagnetic field
\be
	\mathbf{A} = \bar{A}(\bar{r}) \dd t,   \label{eqAphys} 
\ee
which in the Kundt coordinates reads
\be
	\mathbf{A} = A(r) \dd u - \frac{A(r)}{\H(r)} \dd r .
\ee
The Maxwell equations then reduce to 
\be
	A''=0
\ee
and thus in the Kundt coordinates, they can be solved exactly as
\be
A= q r
\ee
(the constant term can be removed by a gauge transformation). 
In contrast, the fundamental electromagnetic invariant $F_{ab} F^{ab}$ can be expressed exactly in the physical coordinates and turns out to be the same as in the Reissner-Nordstr\" om case
\be
F_{ab} F^{ab}=- \frac{2 q^2}{\Omega^4} = - \frac{2 q^2}{{\bar r}^4} \,.
\ee	

Note that the stress-energy tensor \eqref{elTab} and thus the full system of the field equations is invariant under the $SO(2)$ group of electric-magnetic duality. It is thus straightforward to construct a dyonic black hole in quadratic gravity by adding the term $\cal B(\theta) \d \phi$ to the potential \eqref{eqAphys}. Maxwell equations then imply   $\cal B(\theta) = \beta \cos{\theta}$, where $\beta$ is magnetic charge  (the metric remains unchanged except for replacing $q^2$ by $q^2+\beta^2$). This is analogous to the case of the Einstein-Maxwell system, where the dyonic generalization of Reissner-Nordstr\" om black hole admitting both electric and magnetic charges was constructed in \cite{Carter73} (cf also \cite{Ortinbook}).

\subsection{Gravitational field equations}

Following the same approach as in the previous works \cite{Svarcetal18,Podolskyetal18,Podolskyetal20,Pravdaetal21}   for quadratic gravity in vacuum,  we find that the quadratic gravity field equations \eqref{eq:feq}  in the Kundt coordinates \eqref{BHmetric} reduce to 
the following system of  two autonomous 
 ordinary differential equations
\begin{align}
	\Omega\Omega''-2{\Omega'}^2 = &\ \tfrac{1}{3}k\,{\H}'''' \,, \label{Eq1C}\\
	\Omega\Omega'{\H}'+3\Omega'^2{\H}+\Omega^2 %-\Lambda \Omega^4
	= &\ \tfrac{1}{3}k \big({\H}'{\H}'''-{\textstyle\frac{1}{2}}{{\H}''}^2 +2 \big)+ \frac{\kappa'}{2} q^2\,. \label{Eq2C}
\end{align}
Occasionally, it is also useful to use the trace of the field equations which follows from the two above equations
\begin{equation}
	{\H}\Omega''+{\H}'\Omega'+{\textstyle \frac{1}{6}} ({\H}''+2)\Omega =
	0 %{\textstyle \frac{2}{3}\Lambda \,\Omega^3 } 
 \,.
	\label{traceC}
\end{equation}
Note that the only difference from the vacuum case is the term proportional to $q^2$ appearing in 
\eqref{Eq2C}.

\section{Expansion of the metric in powers of $r$ around any fixed finite value $r_0$}
\label{sec_expr0}

In this section, we expand the metric functions $\Omega(r)$ and $\H(r)$ in powers of $\Delta\equiv r-r_0$ around some fixed, finite value $r_0$.  The field equations \eqref{Eq1C}, \eqref{Eq2C} impose constraints on the dominant powers in the expansions, leading to a limited number of allowed classes of static, spherically symmetric power-series solutions of quadratic gravity with electromagnetism summarized in Table \ref{tbl1}.

\subsection{Expansion in powers of~$\Delta$ }
\label{expansio_DElta}

Now, let us solve  \eqref{Eq1C}, \eqref{Eq2C}
 using  expansions in powers of $\Delta$  around an arbitrary, finite, fixed value ${r_0}$,
\begin{eqnarray}
\Omega(r) \rovno \Delta^n   \sum_{i=0}^\infty a_i \,\Delta^{i}\,, \label{rozvojomeg0}\\
\H(r)     \rovno \Delta^p \,\sum_{i=0}^\infty c_i \,\Delta^{i}\,, \label{rozvojcalH0}
\end{eqnarray}
where  $r_0$, $n$, and $p$ are real numbers. We also assume that the leading coefficients are non-vanishing, i.e., $a_0\not=0$, $c_0\not=0$. 
Note that while the steps in ${\Delta}$ are integer
it might not be so in the physical coordinate $\bar r$, see e.g., \cite{Podolskyetal20}.

Substituting the series \eqref{rozvojomeg0}, \eqref{rozvojcalH0}
into the field equations (\ref{Eq1C}), (\ref{Eq2C}), and the trace equation
(\ref{traceC}), yields
\begin{align}
&\sum_{l=2n-2}^{\infty}\Delta^{l}\sum^{l-2n+2}_{i=0}a_i\, a_{l-i-2n+2}\,(l-i-n+2)(l-3i-3n+1) \nonumber \\
& \hspace{45.0mm}=\tfrac{1}{3}k \sum^{\infty}_{l=p-4}\Delta^{l}\,c_{l-p+4}\,(l+4)(l+3)(l+2)(l+1) \,,
\label{KeyEq1C}
\end{align}
\begin{align}
&\sum_{l=2n+p-2}^{\infty}\Delta^{l}\sum^{l-2n-p+2}_{j=0}\sum^{j}_{i=0}a_i\,a_{j-i}\,c_{l-j-2n-p+2}\,(j-i+n)(l-j+3i+n+2) %\nonumber \\
%& \hspace{10.0mm} 
+\sum_{l=2n}^{\infty}\Delta^{l}\sum^{l-2n}_{i=0}a_i\,a_{l-i-2n}
%-\Lambda \sum_{l=4n}^{\infty}
%\Delta^{l}\sum^{l-4n}_{m=0}\bigg(\sum^{m}_{i=0}a_i\,a_{m-%i}\bigg)\bigg(\sum^{l-m-4n}_{j=0}a_j\,a_{l-m-j-4n}\bigg)
\nonumber \\
& = \tfrac{1}{3}k \bigg[2+\sum^{\infty}_{l=2p-4}\Delta^{l}\sum^{l-2p+4}_{i=0}c_{i}\,c_{l-i-2p+4}\,(i+p)(l-i-p+4)
(l-i-p+3)(l-\tfrac{3}{2}i-\tfrac{3}{2}p+\tfrac{5}{2})\bigg]
+\frac{\kappa' q^2}{2}\,,
\label{KeyEq2C}
\end{align}
and 
\begin{align}
&\sum_{l=n+p-2}^{\infty}\Delta^{l}\sum^{l-n-p+2}_{i=0}c_i\,a_{l-i-n-p+2}\,\big[(l-i-p+2)(l+1)
+\tfrac{1}{6}(i+p)(i+p-1)\big] 
%\nonumber \\
%& \hspace{50mm} 
+\tfrac{1}{3}\sum^{\infty}_{l=n}\Delta^{l}\,a_{l-n} =0
%= \tfrac{2}{3}\Lambda
%\sum^{\infty}_{l=3n}\Delta^{l}\sum^{l-3n}_{j=0}\sum^{j}_{i=0}a_i\,a_{j-%i}\,a_{l-j-3n}
\,.
\label{KeyEq3C}
\end{align}
respectively. 

By a careful study of leading orders of equations \eqref{KeyEq1C}-\eqref{KeyEq3C}, it turns out that only certain combinations of exponents $n$, $p$ appearing in the expansion of metric functions $\Omega(r)$ and $\H(r)$, \eqref{rozvojomeg0}, \eqref{rozvojcalH0}, are allowed. The resulting classes $[n,p]$ are summarized in Table \ref{tbl1}.

\begin{table}[H]
		\begin{center}
				\begin{tabular}{|c||c|c|c|}
					\hline
				$[n,p]$	& constraints
    & free parameters & physical region  \\[0.5mm]
					\hline\hline
				$[-1,2]$	& $c_0=-1$	& $a_0$, $c_1$, $r_0$, $q$ &  $\bar r \rightarrow \infty$ \\[1mm]
				$[0,1]$	& 	& $a_0$, $c_0$, $c_1$, $r_0=r_h$, $q$ & $\bar r \rightarrow \bar r_h=a_0 $ \\[1mm]
       $[0,0]$ & &  $a_0$, $a_1$, $c_0$, $c_1$, $c_2$, $r_0$, $q$ & $\bar r \rightarrow \bar r_0=a_0 $  \\[1mm]
       $[0,2] $ & $c_0=-1$, $a_0^2=\frac{\kappa'}{2}q^2$ &  $c_1$, $r_0=r_h$, $q$ & $\bar r \rightarrow \bar r_h=a_0 $ \\[1mm]
       $[1,0]$ & & $a_0$, $c_0$, $c_1$, $c_2$, $r_0$, $q$ & $\bar r \rightarrow 0 $ \\[1mm]
       $[n> 0, 2]$ & $c_0^2 = 1 + 
       \frac{3\kappa'}{4k} q^2$,
       $q^2=-\frac{4kn
       (n+1)[3n(n+1)+2]}{\kappa' [3n(n+1)+1]^2}$, & $a_0$, $c_1$, $r_0$  & $\bar r \rightarrow 0 $ \\
 & $ (3n^2 + 3n + 1)c_0 = -1$ for 
$\frac{ \kappa'}{k} > -\frac{ 4}{3q^2}$  &  & \\
					\hline
				\end{tabular} \\[2mm]
					\caption{All possible classes $[n,p]$ of static, spherically symmetric solutions to quadratic gravity coupled with electromagnetism that can be expressed as power series \eqref{rozvojomeg0}, \eqref{rozvojcalH0}. Note that these classes may admit special subclasses. For example,
 the case $[0,2]$ has special subcases for $c_1=0$, such as $c_2$ arbitrary and $a_0^2=\kappa' q^2/2 = 12 k$, or $c_1=c_2=0$, $c_3$ arbitrary
     and $a_0^2=\kappa' q^2/2 = 24 k$, or in general $c_1=c_2=\dots=c_{l-1}=0$, $c_l$ arbitrary and $a_0^2=\kappa' q^2/2 = 2kl (l+1) $.}
				\label{tbl1}
		\end{center}
\end{table}

Distinct classes [n,p] may or may not correspond to the same branch of solutions in different regions of a spacetime. For example, class [0,1] describing near-horizon metric and class [0,0] with appropriately chosen parameters describing the metric near a generic point outside the horizon may represent the same physical solution. At the same time, class [0,0] with a different choice of parameters can correspond to a generic point outside of a wormhole or naked singularity, see \cite{Podolskyetal20}.

In this paper, we are interested in cases describing black-hole solutions to quadratic gravity in the vicinity of a horizon. Thus in the rest of the paper, we will focus on the classes [0,1] and [0,2], for which $\H=0$ at $r_0=r_h$ \eqref{horizon}, corresponding to non-extremal/extremal horizon, respectively.
Moreover, it seems that the case $[0,2]$ is not asymptotically flat (see section \ref{sec_[0,2]}). Thus, the most physically relevant case is $[0,1]$. 

%degenerate=extremal=double

\section{Charged black holes with non-extremal horizons (case [0,1])}
\label{sec_non-extremal}

So far, in section \ref{sec_expr0}, we  employed the leading orders of equations \eqref{KeyEq1C}-\eqref{KeyEq3C}. Now let us focus on the case corresponding to the solution in the vicinity of a non-extremal horizon, set $[n,p]=[0,1]$, and study conditions following from higher orders in eqs. \eqref{KeyEq1C}-\eqref{KeyEq3C} (or equivalently (\ref{Eq1C}), (\ref{Eq2C}), and (\ref{traceC})).
% This will allow us to find recurrent formulas expressing coefficients $a_l$ and $c_l$ in terms of coefficients with lower indices.  

The lowest nontrivial order of the trace equation \eqref{traceC} gives
\begin{align}
	a_1=-\frac{a_0}{3c_0} (c_1+1)  \,.
	\label{nonSchwinitcond3}
\end{align}
Then the lowest nontrivial order of  \eqref{Eq2C} reduces to 
\be
c_2 =\frac{1}{6kc_0}\left[2k(c_1^2-1)+a_0^2(2-c_1)\right] -\frac{q^2 \kappa'}{4 c_0 k}\,.
\label{nonSchwinitcond2}
\ee
 All higher-order coefficients can be expressed via recurrent formulas
\bea
c_{l+2}\!\!\!&=&\!\!\!\frac{3}{k\,(l+3)(l+2)(l+1)l}\,\sum^{l}_{i=0}a_i \,
a_{l+1-i}(l+1-i)(l-3i)  \qquad \forall\ l\ge 1\,,
\label{nonSchwinitcondc}\\
a_{l}\!\!\!&=&\!\!\!\frac{1}{l^2c_0}\Bigg[
%\tfrac{2}{3}\Lambda \sum^{l-1}_{j=0}{a_{l-1-j}}\sum^{j}_{i=0}a_i\,a_{j-i}
-\tfrac{1}{3}\, a_{l-1}
-\sum^{l}_{i=1}c_i\,a_{l-i}\left[l(l-i)+\tfrac{1}{6}i(i+1)\right]\Bigg] \ \ \
\forall \ l\geq 2\,,  \label{nonSchwinitconda}
\eea
where $a_0$, $c_0$, $c_1$, and $q$ are arbitrary constant parameters. Note that the form of the recurrent expressions \eqref{nonSchwinitcondc}, \eqref{nonSchwinitconda}  is identical to the uncharged case \cite{Podolskyetal20} since charge enters only via $c_2$ in \eqref{nonSchwinitcond2}.

On the horizon, the  Bach and Weyl invariants \eqref{invB}, and \eqref{invC},  read
\bea
B_{ab}\,B^{ab}(r_h) &=& \left(\frac{2 a_0^2 ({c_1}-2)+3 \kappa' 
	q^2
 %+2 a_0^4 \Lambda
 }{12 a_0^4 k}\right)^2  =  \left(\frac{2 a_0^2 b +  \kappa' q^2}{4 a_0^4 k}\right)^2 \,, 
 \label{BachInvariant}\\
 C_{abcd}\,C^{abcd}(r_h) &=&\frac{[2(c_1+1)]^2}{3a_0^4}=3\left(\frac{2(b+1)}{a_0^2}\right)^2\,, 
\eea
respectively,
where, similarly as in the vacuum case, we introduce a {dimensionless} Bach parameter  $b$ measuring the strength of the Bach tensor at the horizon when $q=0$ by
\be
b \equiv \frac{1}{3}\left(c_1-2 %+\Lambda a_0^2
\right)\,. \label{b_definice}
\ee

Note that the invariant \eqref{BachInvariant} vanishes either for $b=0=q$, which corresponds to the Schwarzschild solution, or for $b=-\kappa' q^2/(2a_0^2)$.

Then a few terms in the expansions 
\eqref{rozvojomeg0} and 
\eqref{rozvojcalH0} read
\bea
%\H&=&c_0(r-r_h)+c_1(r-r_h)^2+\frac{2 a_0^2 (2-c_1)  + 4  k(c_1^2-1) - 3 \kappa' q^2 }{ 12k c0 }(r-r_h)^3 %+c_3(r-r_h)^4
%+\dots\,,\nonumber\\
\H &=&
c_0(r-r_h)+(3b+2)(r-r_h)^2+\frac{ 12 kb^2 +2 b( 8 k-a_0^2) +4 k -  \kappa' q^2 }{ 4k c_0 }(r-r_h)^3 %+c_3(r-r_h)^4
+\dots\,,\label{H_[0,1]}\\
%    \Omega&=&a_0-\frac{a_0 (c_1 + 1)}{3 c_0}(r-r_h)
 %   +\frac{a_0}{144 k c_0^2 } \left[16 (c_1 + 1)^2 k + 9\kappa' q^2  + 6 a_0^2 ( c_1 - 2  )\right](r-r_h)^2
      %+a_3(r-r_h)^3+a_4(r-r_h)^4
 %     +\dots\,,\nonumber\\
 \Omega   &=&a_0-\frac{a_0 (b+1)}{ c_0}(r-r_h)
    +\frac{a_0}{16 k c_0^2 } \left[16 (b+1)^2 k  + 2 a_0^2 b+ \kappa' q^2  \right](r-r_h)^2
      %+a_3(r-r_h)^3+a_4(r-r_h)^4
      +\dots\,.\label{Omega_[0,1]}
\eea

Therefore, using \eqref{to static},
%for ${\Delta\equiv r-r_h\to 0}$, 
%\be
%\bar r=\Omega \rightarrow \ a_0 %=-\frac{1}{r_h}}
% \,,  \ \
%\H \rightarrow  c_0\,\Delta
%%=r_h(r-r_h)}
%\,
% \ee
% and thus
in physical coordinates, the horizon is located at\footnote{Note that without loss of generality, the sign of $a_0$ can be chosen thanks to the invariance of \eqref{BHmetric} under the sign change $\Omega\ \rightarrow\ -\Omega$.} 
 \be
\bar{r}(r_h)=\Omega(r_h)=\bar r_h =a_0>0\label{hor_a0} \,,
\ee
where both metric functions $f$ and $h$ vanish  (c.f. \eqref{Schwarz}).

Using \eqref{scalingfreedom}, one can also set $a_1>0$, to ensure monotonic increasing of $\bar r$ as a function of $r$ in the vicinity of the horizon. Finally, for studying solutions around the outer black hole horizon, we assume  $c_0<0$ (see \cite{PraPraOrt23}). To summarize, 
\be
a_0\,,\ a_1>0\,,\quad c_0<0.
\ee
%that can be achieved again using }

%There are two generalizations \\
%$\bullet$ electromagnetic generalization of Schwarzschild =  quadro--electro--%Schwarzschild and \\
%$\bullet$ electromagnetic generalization of Schwa--Bach solution = quadro--%elecro--Schwa--Bach 

%In the second case, the limit $b=0=q$ gives Schwarzschild and we can use the %same gauge as in [], e.g., $c_0=r_h$ and $a_0=-1/r_h$.

It turns out that similarly as in the vacuum case \cite{Luetal15,Luetal15b,Podolskyetal18}, for a generic value of the Bach parameter $b$, the metric \eqref{physmet} expressed using the expansions \eqref{rozvojomeg0}, \eqref{rozvojcalH0} and the recurrent formulas 
\eqref{nonSchwinitcondc} and \eqref{nonSchwinitconda} is not asymptotically flat.
However, for given specific values of parameters, the parameter $b$ can be fine-tuned to obtain asymptotically flat cases. 
Let us thus now present examples of [0,1] solutions for some specific values of parameters.

We will  choose
\be
r_0=-1,\ k=1/2,\ %\tg{\Lambda=0},\ 
c_0=-1, \
a_0 = 1,\ \kappa' = 1,
\label{parameters}
\ee
together with a selected value of charge $q$ and corresponding fine-tuned values of $b$. For $q< \approx0.91$ there will be two such values of $b$. The corresponding two branches of solutions represent charged Schwarzschild\footnote{Recall that charged Schwarzschild in quadratic gravity is distinct from Reissner-Nordstr\" om metric which is not a solution of quadratic gravity with Maxwell field. } and charged Schwarzschild-Bach solutions in quadratic gravity.

\begin{figure}[h!]
    \centering
    \includegraphics[height=60mm]{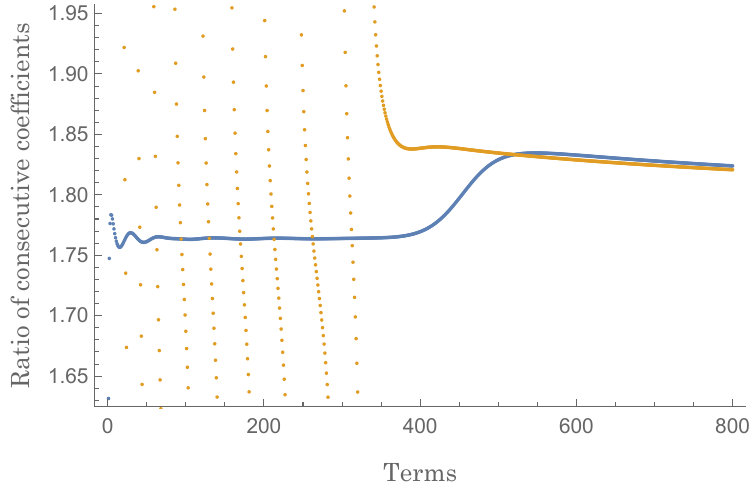}
    \caption{Ratios of consecutive coefficients, $a_i/a_{i-1}$ (blue) and $c_i/c_{i-1}$ (yellow), in the expansions of metric functions \eqref{rozvojomeg0}, \eqref{rozvojcalH0}, using the parameters \eqref{parameters} with $b=0.6314924332806$ and $q=0.5$.}
    \label{fig:terms}
\end{figure}

%\subsection{Specific illustrative example - pictures to be added by George}
First note that often, the series  $a_i$ and $c_i$ asymptotically approach geometric series\footnote{ If the series  $a_i$ and $c_i$ do not asymptotically approach geometric series, more general criteria have to be used for determining convergence of the series. For example, in this context, the root test for convergence is employed in \cite{Ortaggio24prep}.} (see Figure \ref{fig:terms}). This will allow us to estimate the radius of convergence of the series.

%Refer back to these parameters in each caption. Exclude b and q but metion specifically in each plot.

\begin{figure}[h!]
    \centering
    \begin{minipage}{0.48\textwidth}
        \centering
        \includegraphics[width=\linewidth]{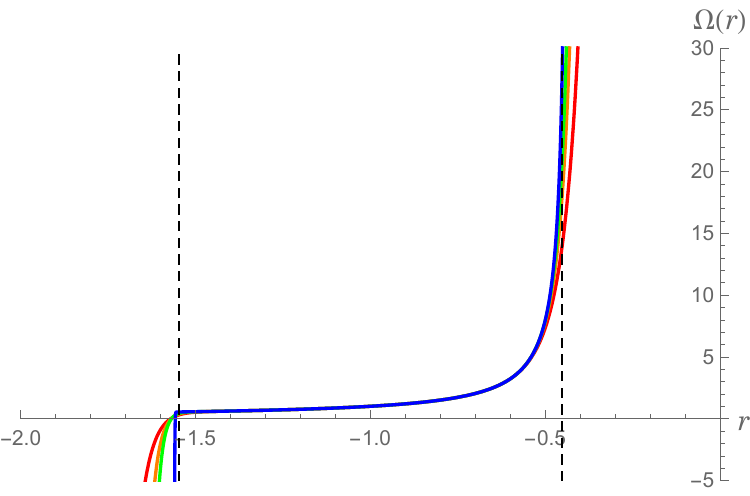}
        \caption{The metric function $\Omega(r)$, using the parameters \eqref{parameters} with $b=0.6314924332806$ and $q=0.5$, for the first 20 (red), 30 (orange), 40 (green), and 800 (blue) terms. The vertical dashed lines indicate the interval of convergence of the series ($r_0\pm 1/p$ where $p$ is the value converged upon in Figure \ref{fig:terms}). }
        \label{fig:omega}
    \end{minipage}%
    \hfill
    \begin{minipage}{0.48\textwidth}
        \centering
        \includegraphics[width=\linewidth]{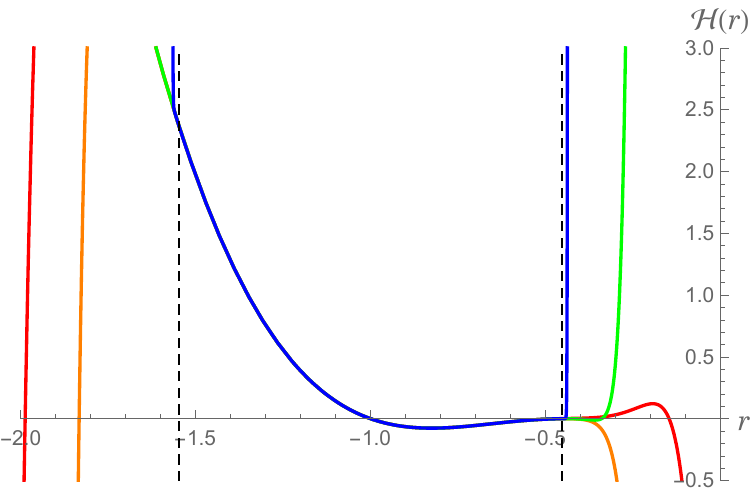}
        \caption{The metric function $\H(r)$, using the parameters \eqref{parameters} with $b=0.6314924332806$ and $q=0.5$, for the first 20 (red), 30 (orange), 40 (green), and 800 (blue) terms. The vertical dashed lines indicate the interval of convergence of the series as in Figure \ref{fig:omega}.}
        \label{fig:scH}
    \end{minipage}
\end{figure}

\begin{figure}[h!]
    \centering
    \begin{minipage}{0.48\textwidth}
        \centering
        \includegraphics[width=\linewidth]{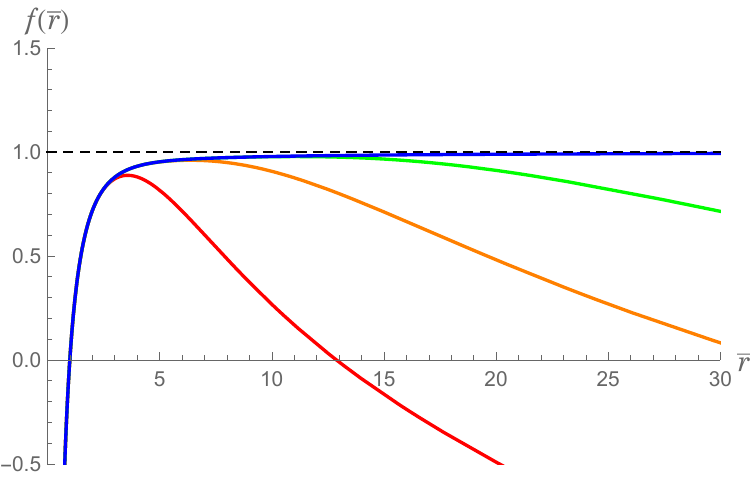}
        \caption{The metric function $f(\bar{r})$, using the parameters \eqref{parameters} with $b=0.6314924332806$ and $q=0.5$, for the first 20 (red), 50 (orange), 100 (green), and 800 (blue) terms.}
        \label{fig:f}
    \end{minipage}%
    \hfill
    \begin{minipage}{0.48\textwidth}
        \centering
        \includegraphics[width=\linewidth]{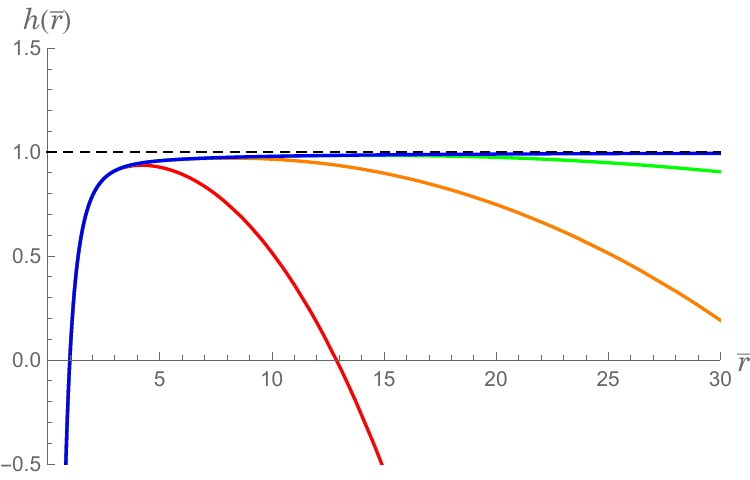}
        \caption{The metric function $h(\bar{r})$, using the parameters \eqref{parameters} with $b=0.6314924332806$ and $q=0.5$, re-scaled to approach 1 at large $\bar{r}$. The first 20 (red), 50 (orange), 100 (green), and 800 (blue) terms of the expansion are shown.}
        \label{fig:h}
    \end{minipage}
\end{figure}

The behavior of the metric functions $\Omega(r)$ and $\H(r)$ for parameters \eqref{parameters}, $q=0.5$, and  $b=0.6314924332806$ is shown in Figures 
\ref{fig:omega} and \ref{fig:scH}, respectively. Using  \eqref{Schwarz}, one can also express the metric functions $f(\bar{r})$ and $h(\bar{r})$ in the physical coordinates, see Figures 
\ref{fig:f} and \ref{fig:h}, respectively. Note that the function $h(\bar{r})$ is re-scaled to approach 1 at large $\bar{r}$. 

\begin{figure}[h!]
    \centering
    \includegraphics[height=50mm]{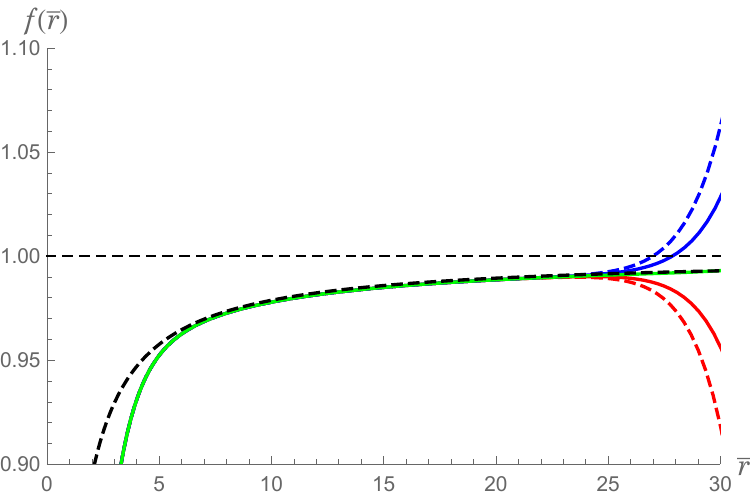}
    \caption{The metric function $f(\bar{r})$ of charged Schwarzschild-Bach solution for various values of $b$ using the parameters \eqref{parameters} and $q=0.5$: $b=0.6314924332804$ (blue, dashed); $b=0.6314924332805$ (blue, solid); $b=0.6314924332807$ (red, solid); $b=0.6314924332808$ (red, dashed); and the most finely-tuned $b=0.6314924332806$ in green. The black, dashed line represents the function $1-2M/\bar{r}$ where $2M \approx 0.212$: the mass determined from the asymptotic behaviour of $h(\bar{r})$.}
    \label{fig:nstuning}
\end{figure}

\begin{figure}[h!]
    \centering
    \includegraphics[height=50mm]{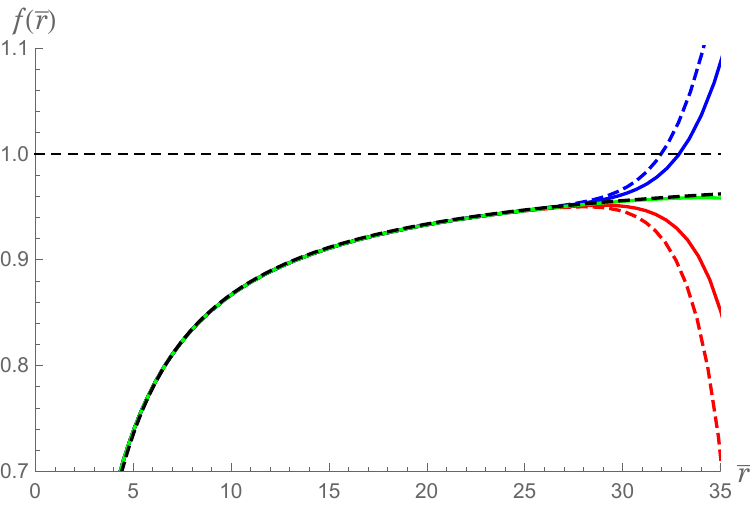}
    \caption{The metric function $f(\bar{r})$ of charged Schwarzschild solution for various values of $b$ using the  parameters \eqref{parameters} and $q=0.5$: $b=-0.269102353931866$ (blue, dashed); $b=-0.269102353931867$ (blue, solid); $b=-0.269102353931869$ (red, solid); $b=-0.26910235393187$ (red, dashed); and the most finely-tuned $b=-0.269102353931868$ in green. The black, dashed line represents the function $1-2M/\bar{r}$ where $2M \approx 1.327$: the mass determined from the asymptotic behaviour of $h(\bar{r})$.}
    \label{fig:stuning}
\end{figure}

The high precision needed for fine-tuning the Bach parameter $b$ is illustrated by Figures \ref{fig:nstuning} and \ref{fig:stuning}. Both figures correspond to the same parameters \eqref{parameters} and $q=0.5$, but are fine-tuned for distinct values of $b$. Figure \ref{fig:nstuning} corresponds to the charged Schwarzschild-Bach black hole while Figure \ref{fig:stuning} corresponds to the charged Schwarzschild black hole.

\begin{figure}[h!]
    \centering
    \includegraphics[height=50mm]{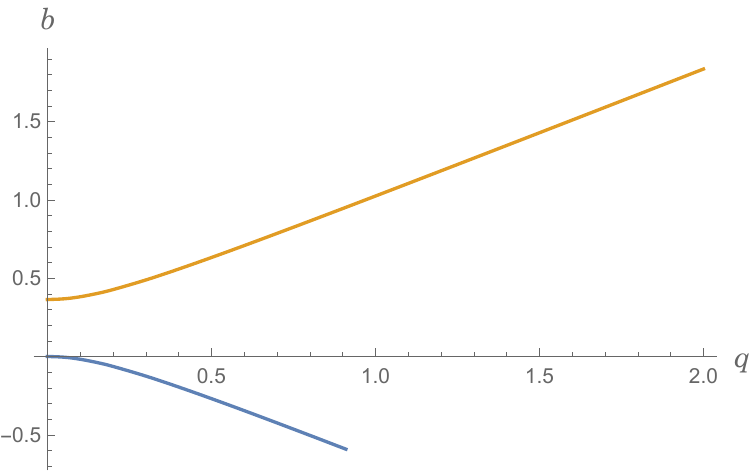}
    \caption{Finely-tuned Bach parameters $b$ corresponding to asymptotically flat solutions (with parameters \eqref{parameters}) as functions of charge $q$. The charged Schwarzschild branch (blue) truncates around $q\approx0.91$ while the charged Schwarzschild-Bach branch (yellow) permits asymptotically flat solutions with greater charge.}
    \label{fig:b}
\end{figure}

These two branches of solutions, charged Schwarzschild and charged Schwarzschild-Bach, are shown in Figure \ref{fig:b}.

%[Please add the following figures (mostly from 01numsoln file) : 1) ratios of the subsequent terms 2)functions Omega and H (color coded according to number of terms), 3) functions f and h (color coded)] 4) show behaviour of f changes with small change of b -> the fine tuning picture

%---------Vojta edited up to here----------------

\subsection{Some physical properties of charged [0,1] black holes}

In this section, we study selected physical properties of charged [0,1] black holes. These include mass, temperature, the radius of the photon sphere, and the black-hole shadow.

Recall that the metric function $h$ of the [0,1] black hole obtained from the recurrent formulas \eqref{nonSchwinitcondc}, \eqref{nonSchwinitconda} with the Bach parameter $b$ fine-tuned for asymptotic flatness and transformed to physical coordinates is in general asymptotically approaching a constant distinct from 1. To set the asymptotic value of $h$ to 1 we employ the scaling freedom \eqref{scaling-t}. This will fix the value of $\sigma$ for each specific asymptotically flat [0,1] black hole. 

\subsubsection{Asymptotic mass, horizon area, surface gravity and temperature}

\begin{figure}[h!]
    \centering
    \includegraphics[height=50mm]{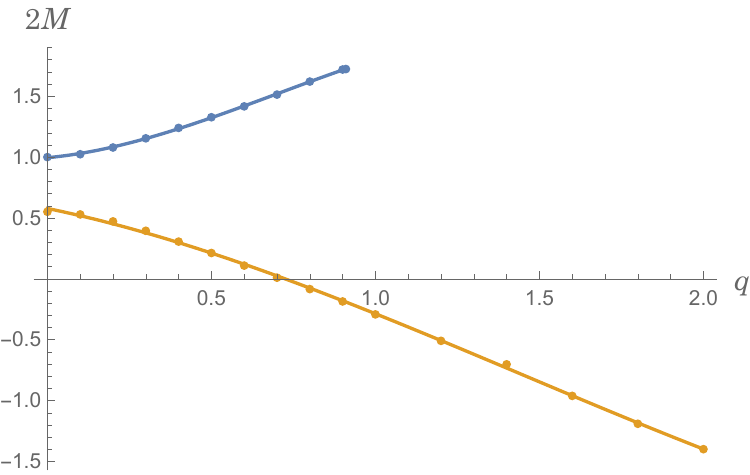}
    \caption{Asymptotic mass as a function of charge $q$, determined by fitting the model $1-2M/\bar{r}$ to the re-scaled metric function $h(\bar{r})$ (see Figure \ref{fig:h}) over an appropriate range of $\bar{r}$. Charged Schwarzschild solutions are in blue while charged Schwarzschild-Bach solutions are in yellow. All solutions make use of the parameters \eqref{parameters}.}
    \label{fig:2M}
\end{figure}

\begin{figure}[h!]
    \centering
    \includegraphics[height=50mm]{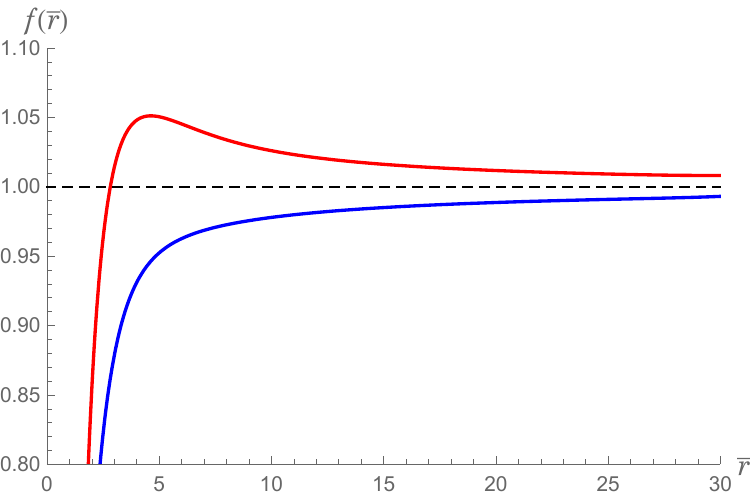}
    \caption{$f(\bar{r})$ for two charged Schwarzschild-Bach solutions using the parameters \eqref{parameters}: one with positive mass, $q=0.5$ and tuned to $b=0.6314924332806$ (blue); the other with negative mass, $q=1$ and tuned to $b=1.02395913547716$ (red).}
    \label{fig:masscomp}
\end{figure}

Asymptotic mass $M$ can be determined by comparing the metric function $f(\bar{r}) $ or the re-scaled metric function $h(\bar{r})$ with  $1-2M/\bar{r}$. 
See Figure \ref{fig:2M} for the dependence of asymptotic mass $M$ on charge $q$ for both branches of asymptotically flat $[0,1]$ black holes.
    Note that while for the Schwarzschild branch, asymptotic mass increases with charge, for  Schwarzschild-Bach it decreases and reaches negative values for sufficiently large charge.\footnote{Note, however, that for sufficiently small $a_0$, this behavior is interchanged, e.g., for the Schwarzschild branch, the asymptotic mass decreases with increasing charge, see \cite{Wu20}.} This can be also seen in Figure \ref{fig:masscomp}. 

\begin{figure}[h!]
    \centering
    \begin{minipage}{0.48\textwidth}
        \centering
        \includegraphics[width=\linewidth]{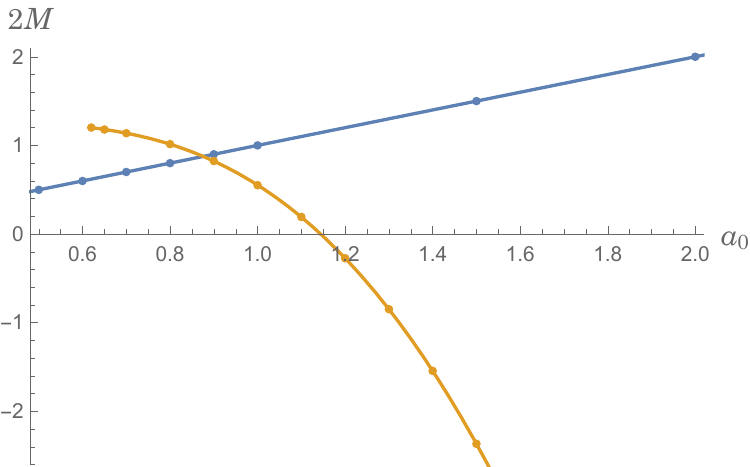}
        \caption{Asymptotic mass of Schwarzschild (blue) and uncharged Schwarzschild-Bach (yellow) black holes as functions of the horizon radius $\bar r_h=a_0$, cf. \eqref{hor_a0}. This figure corresponds to previously published Figure 3 of \cite{Luetal15b}.}
        \label{fig:2Ma00}
    \end{minipage}%
    \hfill
    \begin{minipage}{0.48\textwidth}
        \centering
        \includegraphics[width=\linewidth]{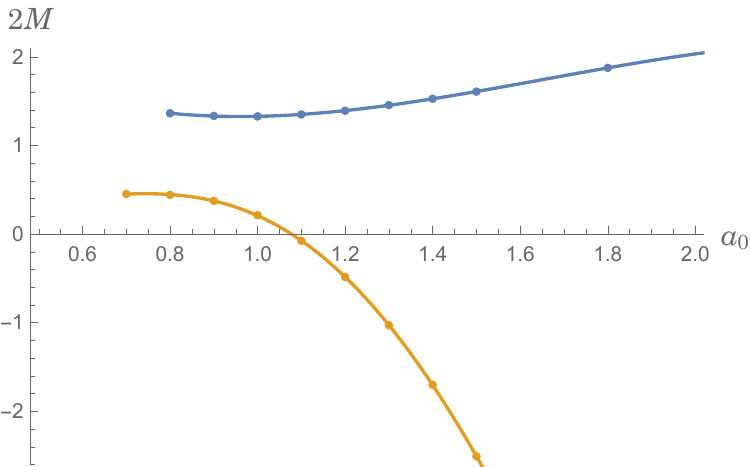}
        \caption{Asymptotic mass of charged Schwarzschild (blue) and Schwarzschild-Bach (yellow) black holes with charge $q=0.5$ as functions of the horizon radius $\bar r_h=a_0$, cf. \eqref{hor_a0}.}
        \label{fig:2Ma05}
    \end{minipage}
\end{figure}

In most figures, we keep the horizon radius $\bar r_h=a_0$, cf. \eqref{hor_a0},  fixed by \eqref{parameters}. 
However, it is also of interest to study the dependence of the asymptotic mass $M$ on the horizon radius $\bar r_h=a_0$,
see Figures \ref{fig:2Ma00}  and \ref{fig:2Ma05} for the uncharged and charged case, respectively. 

%\subsubsection{Area, surface gravity, and temperature}

The null Killing vector ${\xi\equiv\sigma\partial_u=\sigma\partial_t}$ generates the horizon. The location of which is given by vanishing of the norm of $\xi$, i.e., by ${\H(r)=0}$ at ${r=r_h}$,   using
\eqref{H_[0,1]}. The horizon area reads 
\be
{\cal A} = 4\pi\,\Omega^2(r_h)= 4\pi a_0^2 %= \frac{4\pi}{r_h^2}
= 4\pi\,{\bar r}_h^2 \,.
\label{horizon_area}
\ee

\begin{figure}[h!]
    \centering
    \includegraphics[height=60mm]{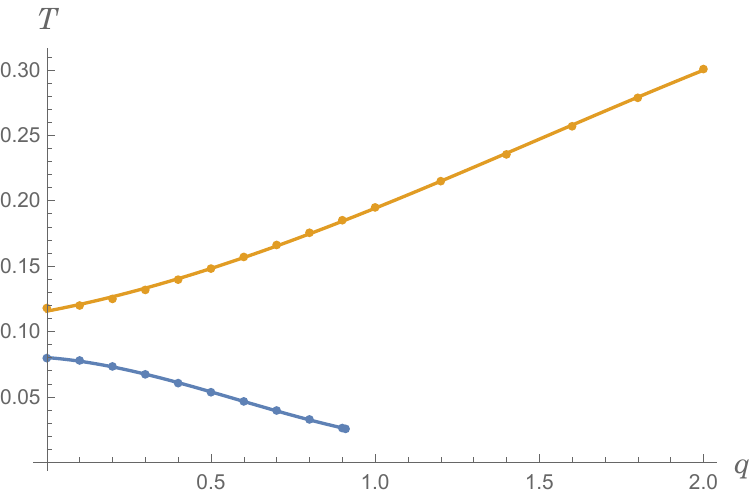}
    \caption{Temperature as a function of charge $q$ for charged Schwarzschild solutions (blue) and charged Schwarzschild-Bach solutions (yellow). All solutions make use of the parameters \eqref{parameters}.}
    \label{fig:T}
\end{figure}

The surface gravity
${\kappa^2\equiv-\frac{1}{2}\,\xi_{\mu;\nu}\,\xi^{\,\mu;\nu}}$
\cite{Waldbook}, where the only nonvanishing derivatives of  $\xi$ are $\xi_{u;r}\! =\!-\xi_{r;u}\! =\!\frac{1}{2}\sigma(\Omega^2\H)'$,
${\xi^{\,r;u}\!=\!-\xi^{\,u;r}\!= \!\Omega^{-4}\xi_{u;r}}$, using \eqref{H_[0,1]},  reads
\be
\kappa/\sigma =-\frac{1}{2}(\H'+2\H\,\Omega'/\Omega)_{|r=r_h} = -\frac{1}{2}\,\H'(r_h) =
-\frac{1}{2}\, c_0 
%\tr{= -\frac{r_h}{2} =\frac{1}{2\, {\bar r}_h}}
\,.
\label{surface_gravity}
\ee	
Then the temperature  of
the black hole horizon ${T\equiv\kappa/(2\pi)}$  \cite{FanLu15} is given by
\be
T/\sigma
= -\frac{1}{4\pi}c_0
%=\tr{-\frac{1}{4\pi} \,r_h
 %       = \frac{1}{4\pi}\,{\bar r}_h^{\,-1} 
 \,.
\label{temperature}
\ee
Figure \ref{fig:T} shows the dependence of temperature on charge for both branches of black holes.

%\tb{The limit $T\rightarrow\ 0$, i.e., $c_0\rightarrow\ 0$, corresponds to the extremal BH with double horizon, i.e., case $[0,2]$. When performing/considering this limit, the ``Schwarzschild'' gauge cannot be used.}

%\tr{[same gauge as in non-charged case? $a_0 =-\frac{1}{r_h}$, $c_0 =r_h$]}

%\tr{[compare with the paper on charged BHs]}

%\tr{[extreme limit? $c_0\rightarrow 0$]}

\subsubsection{Photon sphere and black hole shadow}
\label{sec_photon}

In this section, we study photon spheres and black-hole shadows for asymptotically flat, spherically symmetric, charged black holes in quadratic gravity.

\begin{figure}[h!]
    \centering
    \includegraphics[height=70mm]{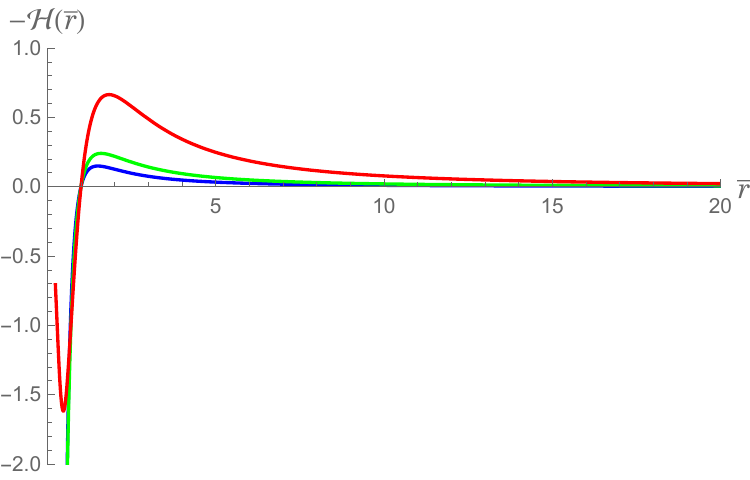}
    \caption{$-\H(\bar{r})$ for various charged Schwarzschild black holes using the parameters \eqref{parameters}: blue ($q=0$, $b=0$), green ($q=0.5$, $b=-0.269102353931868$), and red ($q=0.9$, $b=-0.583718480513498$). This functions as an effective potential for null geodesics around the black hole.}
    \label{fig:schwpss}
\end{figure}

\begin{figure}[h!]
    \centering
    \includegraphics[height=70mm]{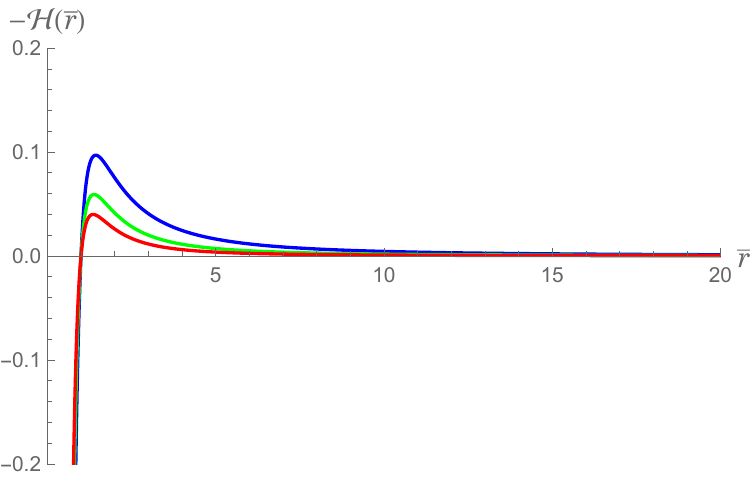}
    \caption{$-\H(\bar{r})$ for various charged Schwarzschild-Bach black holes using the parameters \eqref{parameters}: blue ($q=0$, $b=0.3633018769168$), green ($q=1$, $b=1.02395913547716$), and red ($q=2$, $b=1.8348341584421$). This functions as an effective potential for null geodesics around the black hole.}
    \label{fig:nschwpss}
\end{figure}

In the physical coordinates \eqref{physmet}, the photon sphere\footnote{ For the definition of  photon spheres for static, spherically symmetric spacetimes in standard coordinates \eqref{physmet}, see \cite{ClaVirEll01}.} is located at $2h=\bar r h,_{\bar r}$, i.e., 
\be
\left(\frac{h}{\bar r^2}\right),_{\bar r}=0\,.\label{ph_sphere_h}
\ee
As pointed out in \cite{Ortaggio24prep}, the Kundt coordinates \eqref{BHmetric} are more suitable for the study of photon spheres. In these coordinates, using \eqref{Schwarz}, \eqref{ph_sphere_h} reduces to \cite{Ortaggio24prep}
\be
\H,_r (r_{\rm ps})=0\,,
\ee
 provided ${\H}(r_{\rm ps})<0$ and  $\Omega,_r (r_{\rm ps})\not=0$. 
Photon spheres in fact correspond to local maxima of the metric function $-\H$,
see Figures \ref{fig:schwpss} and  \ref{fig:nschwpss} for
charged Schwarzschild and charged Schwarzschild-Bach black holes with different charges, respectively.

\begin{figure}[h!]
    \centering
    \includegraphics[height=70mm]{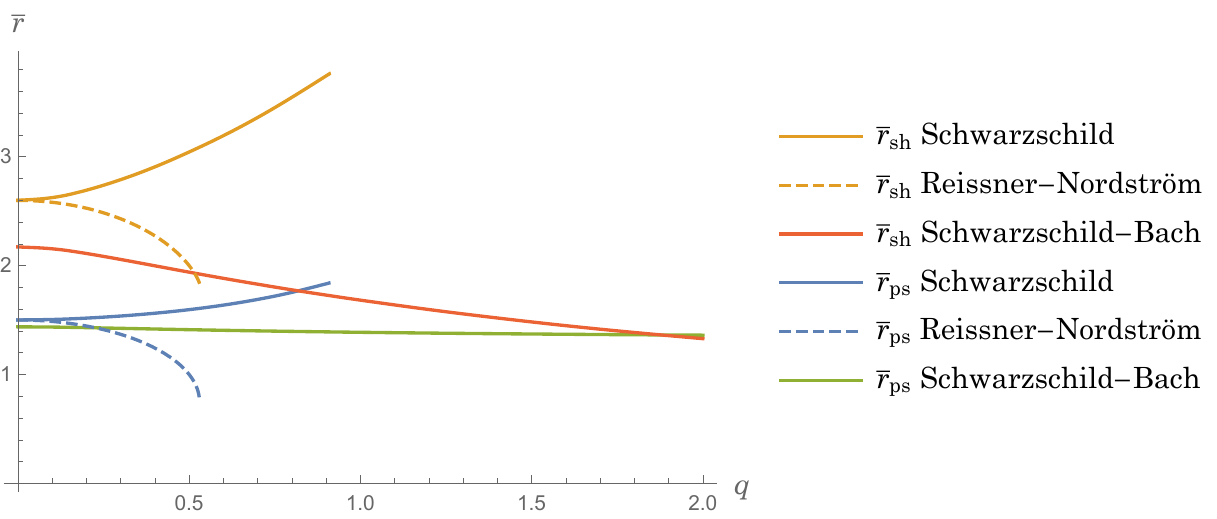}
    \caption{Radii $\bar{r}$ of photon spheres and black hole shadows as functions of charge $q$ and using the parameters \eqref{parameters}. The photon spheres of charged Schwarzschild and Schwarzschild-Bach black holes correspond to the solid blue and green lines respectively; similarly, the solid yellow and red lines correspond to the respective shadows. The dashed lines correspond to the Reissner-Nordstr\" om black hole in contrast to the charged Schwarzschild solutions of quadratic gravity.}
    \label{fig:psshadow}
\end{figure}

\begin{figure}[h!]
    \centering
    \begin{minipage}{0.48\textwidth}
        \centering
        \includegraphics[width=\linewidth]{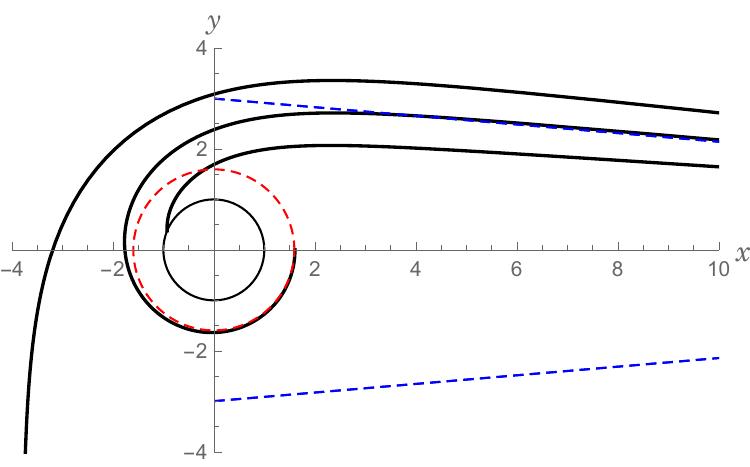}
        \caption{Null geodesics in the vicinity of a charged Schwarzschild black hole with the following parameters: $a_0=1$, $q=0.5$, $b=-0.269102353931868$. Each passes through a distant point ($\bar{r}\approx35$) where the fine-tuning of $b$ is sufficient to maintain asymptotic-flatness. From this point, the projection of the angular shadow is indicated by the dashed, blue lines. The black circle and dashed, red circle indicate the event horizon and photon sphere, respectively.}
        \label{fig:pmorbit}
    \end{minipage}%
    \hfill
    \begin{minipage}{0.48\textwidth}
        \centering
        \includegraphics[width=\linewidth]{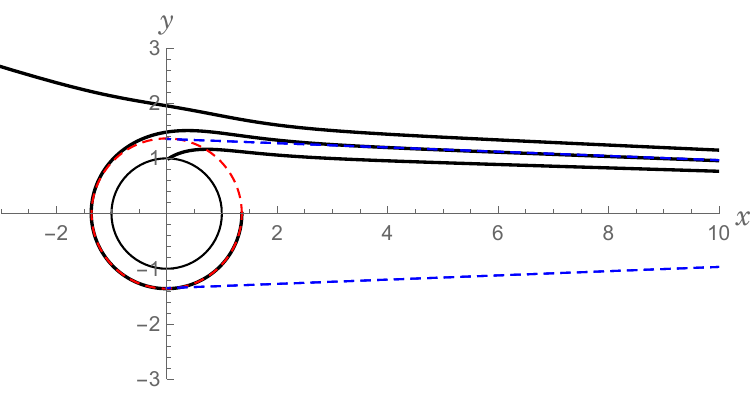}
        \caption{Null geodesics in the vicinity of a charged Schwarzschild-Bach black hole with the following parameters: $a_0=1$, $q=2$, $b=1.8348341584421$. Each passes through a distant point ($\bar{r}\approx35$) where the fine-tuning of $b$ is sufficient to maintain asymptotic flatness. From this point, the projection of the angular shadow is indicated by the dashed, blue lines. The black circle and dashed, red circle indicate the event horizon and photon sphere, respectively. It should be noted that the negative lensing of light not captured by the black hole is due to the negative mass of the highly charged Schwarzschild-Bach black hole.}
        \label{fig:nmorbit}
    \end{minipage}
\end{figure}

The shadow of an asymptotically flat, spherically symmetric black hole is a disk of a radius $\bar r_{\rm sh}=\frac{\bar r_{\rm ps}}{\sqrt{h(\bar r_{\rm ps})}}$ (see (2.14) of \cite{Lu2020}). In the Kundt coordinates, this reads
\be
\bar r_{\rm sh}
=\frac{1}{\sigma\sqrt{|\H(r_{\rm ps})|}}\,.\label{rsh}
\ee
See Figure \ref{fig:psshadow}, for the comparison of photon-sphere and black-hole-shadow radii, $\bar r_{\rm ps}$ and $\bar r_{\rm sh}$,  for various charged Schwarzschild and Schwarzschild-Bach black holes.
Note that for highly charged Schwarzschild-Bach black holes, $\bar r_{\rm sh}<\bar r_{\rm ps} $. This is due to the negative lensing of light caused by negative mass of these black holes (see also Figure \ref{fig:nmorbit}).

\begin{figure}[h!]
    \centering
    \includegraphics[height=35mm]{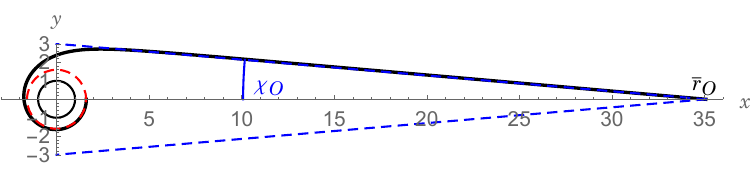}
    \caption{A critical null geodesic entering the photon sphere of a charged Schwarzschild black hole with the following parameters: $a_0=1$, $q=0.5$, $b=-0.269102353931868$. The black circle and dashed, red circle indicate the event horizon and photon sphere respectively. An observer situated at $\bar{r}_O$ will measure $\chi_O$ as the angular size of the black hole.}
    \label{fig:pmorbitfull}
\end{figure}

Finally, let us discuss the angular radius $\chi_O$ of the black-hole shadow (see Figure \ref{fig:pmorbitfull} and review \cite{Perlick2022}).
Following  \cite{Perlick2015}, the angular radius of the black-hole shadow for a static observer at $\bar r=\bar r_O$ in the static exterior region reads
\be
\cot{\chi_O}=\sqrt{\frac{g_{\bar r \bar r}}{g_{\phi\phi}}}
\frac{\d \bar r}{\d \phi}_{| \bar r=\bar r_O}
=\frac{1}{\bar r\sqrt{f}}\frac{\d \bar r}{\d \phi}_{| \bar r=\bar r_O}\,,
\quad 
%where $
\frac{\d \bar r}{\d \phi}=\pm \bar r\sqrt{f}\sqrt{\frac{\bar r^2 h(\bar r_{\rm ps})}{\bar r_{\rm ps}^2 h(\bar r)}-1}\,.
\ee
Thus 
\be
\sin^2 \chi_O=\frac{1}{1+\cot^2 \chi_O}=\frac{\bar r_{\rm ps}^2 h(\bar r_O)}{\bar r_{O}^2h(\bar r_{\rm ps})}\,,
\ee
%$ 
which  after using \eqref{Schwarz}, simplifies to (see \cite{Ortaggio24prep})\footnote{Note that in the limit $\bar r_O\ \rightarrow\ \infty$ (then $\sin^2\chi_O \approx \frac{\bar r^2_{sh}}{\bar r^2_O}$) using the gauge such that $h(\bar r_O)=\bar r^2_O\H (r_O)\ \rightarrow\ 1$, \eqref{chiO} leads to \eqref{rsh}.}
\be
\sin^2 \chi_O=\frac{\H (r_O)}{\H (r_{\rm ps})}\,.\label{chiO}
\ee

See Figures \ref{fig:pmorbit} and \ref{fig:nmorbit} for numerically integrated null geodesics for black holes in quadratic gravity with positive and negative mass, respectively.

\begin{figure}[h!]
    \centering
    \begin{minipage}{0.48\textwidth}
        \centering
        \includegraphics[width=\linewidth]{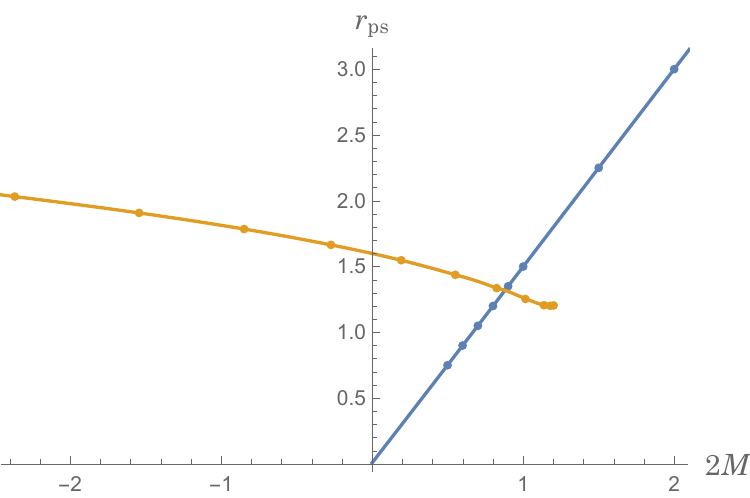}
        \caption{Radii of photon spheres for Schwarzschild (blue) and uncharged Schwarzschild-Bach (yellow) black holes as functions of asymptotic mass.}
        \label{fig:psM0}
    \end{minipage}%
    \hfill
    \begin{minipage}{0.48\textwidth}
        \centering
        \includegraphics[width=\linewidth]{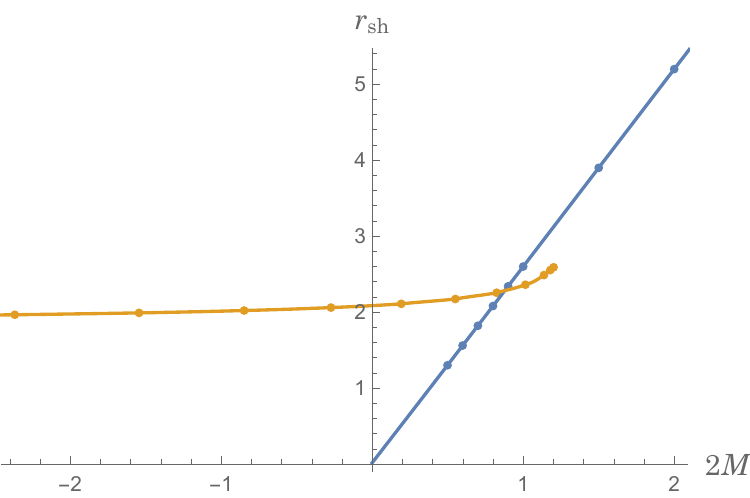}
        \caption{Radii of black hole shadows for Schwarzschild (blue) and uncharged Schwarzschild-Bach (yellow) black holes as functions of asymptotic mass.}
        \label{fig:shM0}
    \end{minipage}
\end{figure}

\begin{figure}[h!]
    \centering
    \begin{minipage}{0.48\textwidth}
        \centering
        \includegraphics[width=\linewidth]{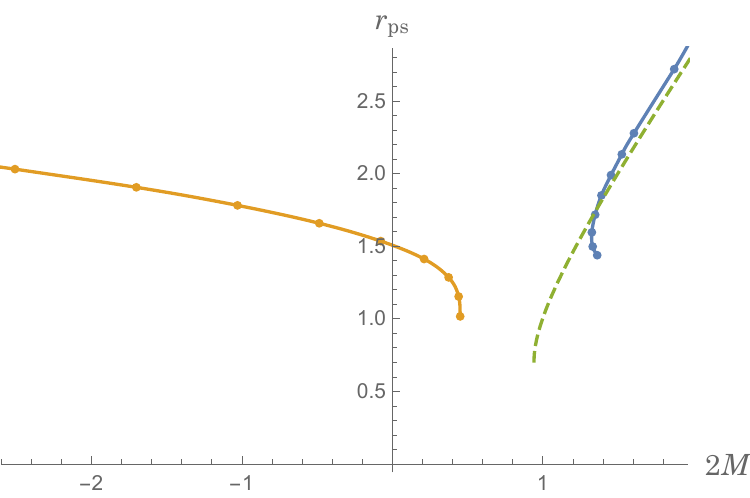}
        \caption{Radii of photon spheres for charged Schwarzschild (blue) and Schwarzschild-Bach (yellow) black holes with charge $q=0.5$ as functions of asymptotic mass. The radii of Reissner-Nordstr\" om black hole photon spheres are indicated by the dashed green line.}
        \label{fig:psM5}
    \end{minipage}%
    \hfill
    \begin{minipage}{0.48\textwidth}
        \centering
        \includegraphics[width=\linewidth]{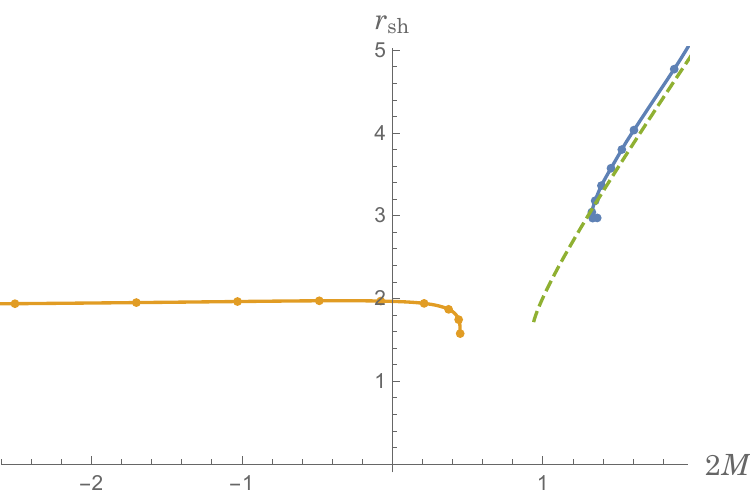}
        \caption{Radii of black hole shadows for charged Schwarzschild (blue) and Schwarzschild-Bach (yellow) black holes with charge $q=0.5$ as functions of asymptotic mass. The radii of Reissner-Nordstr\" om black hole shadows are indicated by the dashed green line.}
        \label{fig:shM5}
    \end{minipage}
\end{figure}

Finally, let us show the dependence of the photon-sphere radius and black-hole shadow radius on asymptotic mass for uncharged (Figures \ref{fig:psM0} and \ref{fig:shM0}) and charged (Figures \ref{fig:psM5} and \ref{fig:shM5})
Schwarzschild and Schwarzschild-Bach black holes.

%\begin{figure}[h!]
 %   \centering
  %  \includegraphics[height=35mm]{pmorbit3.pdf}
   % \caption{A critical null geodesic entering the photon sphere of a charged Schwarzschild black hole with the following parameters: $a_0=1$, $q=0.5$, $b=-0.269102353931868$. The black circle and dashed, red circle indicate the event horizon and photon sphere respectively. An observer situated at $\bar{r}_O$ will measure $\chi_O$ as the angular size of the black hole.}
    %\label{fig:pmorbitfull}
%\end{figure}

% Insert figures for the branches - dependence of b, M and T on q (as in the file 01results) 

 \section{Extremal case $[0,2]$}
 \label{sec_[0,2]}

Finally, let us briefly comment on the case $[0,2]$ describing geometry in the vicinity of the horizon of an extremal black hole.

Eq. \eqref{KeyEq3C} for $l=0$ gives
\be
c_0=-1. \label{02c0}
\ee
Then eq. \eqref{KeyEq2C} for $l=0$ implies
\be 
a_0^2=\frac{\kappa'}{2} q^2\,. \label{02a0}
\ee 
Thus this case can occur only for $q\not=0$. This is in agreement with the results of \cite{Podolskyetal18,Podolskyetal20} in vacuum, where the [0,2] case is not allowed.

Eq. \eqref{KeyEq2C} for $l=1$ implies
\be 
a_1=\frac{a_0 c_1}{2},
\ee 
where $c_1$ is an integration constant. Therefore, [0,2] is characterized by just 2 constants:  a Bach parameter $b_2\equiv c_1$ and charge $q$.

Then combining \eqref{KeyEq1C} and \eqref{KeyEq2C}, for $l\geq 2$, we arrive at the recurrent relations
\bea
c_l&=&-6
\frac{(l+1){\cal Y}_l + a_0 (l-1) {\cal X}_l}{(l-1)(l+2)(l+1)\left[a_0^2-2kl(l+1)\right]}\,,\label{[0,2]cl}\\
a_l&=&-\frac{ a_0 {\cal Y}_l +2 k l (l-1) {\cal X}_l}{l(l-1)\left[ a_0^2-2kl(l+1)\right]}\,,\label{[0,2]al}
\eea 
where
\bea 
{\cal X}_l&=&\sum^{l-1}_{i=1}c_i a_{l-i} \left[(l-i) (l+1)+\frac{1}{6}(i+2)(i+1)\right]\,, \label{[0,2]Xl}\\
{\cal Y}_l&=&\sum^{l-1}_{i=1}a_i a_{l-i} (l-i) (l-1-3i)\,.\label{[0,2]Yl}
\eea 

Near the extremal horizon,  the  Bach and Weyl invariants  \eqref{invB}, \eqref{invC} take the form 
\bea
B_{ab}{ B^{ab}}_{|r=r_h} &=&
%\frac{9 c_1^4 }{2a_0^4 (a_0^2-12 k)^2} \Delta^4
%+{\cal O} (\Delta^5)=
%\frac{72b_2^4 }{\kappa'^2 q^4 (\kappa' q^2 -24 k)^2}\Delta^4+{\cal O} (\Delta^5)\nonumber\\
%&& =
2\left(\frac{6}{\kappa'(\kappa' q^2-24 k)}\right)^2\left(\frac{b_2 \Delta}{q}\right)^4+{\cal O} (\Delta^5)\,,\label{[0,2]BB}\\
C_{abcd}{ C^{abcd}}_{|r=r_h} &=&
%\frac{12 c_1^2 }{a_0^4}\Delta^2+{\cal O} (\Delta^3)=
\frac{48 b_2^2 }{\kappa'^2 q^4} \Delta^2
+{\cal O} (\Delta^3)\,,\label{[0,2]CC}\\
\eea
where $\Delta=r-r_h$ with $r_h$ being the extremal horizon. Thus, similarly as for the Reissner-Nordstr\" om metric and in contrast with the [0,1] case, both the Bach and Weyl invariants vanish on the horizon.

We were not able to fine-tune the $[0,2]$ case to obtain an asymptotically flat solution. This might be related to the fact that the $[0,2]$ solution has less free parameters than the $[0,1]$ black holes.

\subsection{Special cases with $c_1=0$ }

Note that the recurrent expressions \eqref{[0,2]cl}--\eqref{[0,2]Yl}
do not include special cases for which, in addition to \eqref{02c0}, $c_1=0$ and $a_0^2=\frac{\kappa'}{2} q^2=2kl(l+1)$
%, e.g., $a_0^2=12k$, $a_0^2=24k$, $a_0^2=40k$, 
with $c_p=0$, $p=1,\dots, l-1$, and the only non-vanishing coefficients are $c_l\not=0$ ($l\geq 2$), $c_{lm}$, and $a_{lm}$, $m=1,2\dots$.
The coefficients $c_{lm}$ and $a_{lm}$ are proportional to  $({c_l})^m$.
%\tb{For $l=2$, i.e., $a_0^2=12 k$, $c_2\not=0$ and $c_{2m},\ a_{2m}\propto 
%{c_2}^m$, $m=1,2\dots$.} \tb{For $l=3$, i.e., $a_0^2=24 k$, $c_3\not=0$ and $c_{3m},\ a_{3m}\propto {c_3}^m$, $m=1,2\dots$.}

In contrast with the generic $[0,2]$ solution, 
%\eqref{[0,2]BB}, \eqref{[0,2]CC}
near the horizon, the Bach and Weyl invariants  \eqref{invB}, \eqref{invC} go to zero as
\bea
{B_{ab} B^{ab}}_{|r=r_h}&\propto &  \frac{c_l^2}{a_0^8}\Delta^{2l}\,,\\
{C_{abcd}C^{abcd}}_{|r=r_h}&\propto & \frac{c_l^2}{a_0^4}\Delta^{2l}
\,.
\eea

In physical coordinates, the power-series expansions of metric functions $f(\bar r)$ and $h(\bar r)$ \eqref{Schwarz} %for the above special cases
include fractional powers of $\bar\Delta=\bar r-a_0$   since  $\bar r=\Omega=a_0+a_{l}\Delta^l+\dots$ \eqref{to static} and therefore
$\Delta\propto \left(\frac{\bar\Delta}{a_l}\right)^{1/l}$.

\section*{Acknowledgements}

We thank Marcello Ortaggio for helpful comments. This work has been supported by the Institute of Mathematics, Czech Academy of Sciences (RVO 67985840).

\newpage

\appendix
%\appendixpage
\section{Bach parameters required for asymptotic flatness}

\begin{footnotesize}
\begin{longtable}{|c|c|c|c|}
    \hline
    $q$ & $a_0$ & $b$ (un)charged Schwarzschild & $b$ (un)charged Schwarzschild-Bach \\[0.5mm]
    \hline\hline
    0 & 0.62 & 0 & -0.589993571 \\[1mm]
    0 & 0.65 & 0 & -0.532684823 \\[1mm]
    0 & 0.7 & 0 & -0.42996068 \\[1mm]
    0 & 0.8 & 0 & -0.19860582 \\[1mm]
    0 & 0.9 & 0 & 0.0660741 \\[1mm]
    0 & 1 & 0 & 0.3633018769168 \\[1mm]
    0 & 1.1 & 0 & 0.69267676 \\[1mm]
    0 & 1.2 & 0 & 1.0539754 \\[1mm]
    0 & 1.3 & 0 & 1.4470654 \\[1mm]
    0 & 1.4 & 0 & 1.8718642 \\[1mm]
    0 & 1.5 & 0 & 2.3283181 \\[1mm]
    0.05 & 1 & -0.00462 & 0.36791 \\[1mm]
    0.1 & 1 & -0.017831007 & 0.381097084 \\[1mm]
    0.15 & 1 & -0.03810 & 0.40132 \\[1mm]
    0.2 & 1 & -0.063682753 & 0.426841089 \\[1mm]
    0.3 & 1 & -0.1252725399 & 0.488250191 \\[1mm]
    0.4 & 1 & -0.19498743703 & 0.557709781 \\[1mm]
    0.5 & 0.7 &  & 0.243104319 \\[1mm]
    0.5 & 0.8 & -0.5171222483 & 0.320022494 \\[1mm]
    0.5 & 0.9 & -0.3780781576 & 0.444080714 \\[1mm]
    0.5 & 1 & -0.269102353931868 & 0.6314924332806 \\[1mm]
    0.5 & 1.1 & -0.1939706824 & 0.885277714 \\[1mm]
    0.5 & 1.2 & -0.1447671171 & 1.19712757 \\[1mm]
    0.5 & 1.3 & -0.1121881706 & 1.5575176 \\[1mm]
    0.5 & 1.4 & -0.0898183758 & 1.95990043 \\[1mm]
    0.5 & 1.5 & -0.0738318916 & 2.40036761 \\[1mm]
    0.5 & 1.8 & -0.04584801196 &  \\[1mm]
    0.6 & 1 & -0.345793969 & 0.7077718847 \\[1mm]
    0.7 & 1 & -0.4241119294 & 0.7855943886 \\[1mm]
    0.8 & 1 & -0.503525436775 & 0.86442570416 \\[1mm]
    0.85 & 1 & -0.54353900000085 &  \\[1mm]
    0.9 & 1 & -0.583718480513498 & 0.94394733295808 \\[1mm]
    0.91 & 1 & -0.59177226034189 &  \\[1mm]
    1 & 1 &  & 1.02395913547716 \\[1mm]
    1.2 & 1 &  & 1.18496948744475 \\[1mm]
    1.4 & 1 &  & 1.34682262993053 \\[1mm]
    1.6 & 1 &  & 1.5091937242 \\[1mm]
    1.8 & 1 &  & 1.6719006501 \\[1mm]
    2 & 1 &  & 1.8348341584421 \\[1mm]
    \hline
    \caption{Fine-tuned Bach parameters $b$ resulting in asymptotic flatness for a range of charged and uncharged Schwarzschild and Schwarzschild-Bach solutions. These values reflect the extent of fine-tuning carried out in this study and each value can undergo further tuning to achieve flatness at greater distances.}
    \label{tbl2}
\end{longtable}
\end{footnotesize}

	%\bibliographystyle{unsrt}
	%\bibliography{bibl,biblqg_mod}

	\end{document}